\documentstyle[12pt]{article}
\advance\voffset by -0.8in
\advance\hoffset by -0.55in
\begin{document}
\newcommand{\reg}[1]{(\ref{#1})}
\newcommand{\sqr}[2]{{{\vcenter{\vbox{\hrule height.#2pt
\hbox{\vrule width.#2pt height#1pt \kern#1pt
\vrule width.#2pt}
\hrule height.#2pt}}}}}
\renewcommand{\square}{{\mathchoice\sqr{17}4\sqr{17}4\sqr{13}3\sqr{13}3}}
\newcommand{\sgm}{{\rm sign}(m)}
\newcommand{\sgn}{{\rm sign}(n)}
\newcommand{\Hil}{{\cal H}}
\newcommand{\C}{{\cal C}}
\newcommand{\ze}{{\bf Z}}
\newcommand{\re}{{\bf R}}
\newcommand{\ce}{{\bf C}}
\newcommand{\Aut}{{\rm Aut}}
\renewcommand{\C}{{\cal C}}
\title{Orbifold Constructions and the Classification
of Self-Dual $c=24$ Conformal Field Theories}
\author{P.S. Montague\\
Department of Applied Mathematics and Theoretical Physics\\
University of Cambridge\\
Silver Street\\
Cambridge CB3 9EW\\
U.K.}
\maketitle
\begin{abstract}
We discuss questions arising from the work of
Schellekens\cite{Schell:Venkov,SchellComplete}.
After introducing
the concept of complementary representations, we examine $\ze_2$-orbifold
constructions in general,
and propose a technique for identifying the orbifold theory without
knowledge of its explicit construction. This technique is then
generalised to twists of order 3, 5 and 7, and we proceed to apply our
considerations to the FKS constructions $\Hil(\Lambda)$ ($\Lambda$ an
even self-dual lattice) and the reflection-twisted orbifold theories
$\widetilde\Hil(\Lambda)$, which together remain the only $c=24$ theories which
have so far been proven to exist\cite{DGMtwisted}. We also make, in
the course of our arguments, some comments on the automorphism groups
of the theories $\Hil(\Lambda)$ and $\widetilde\Hil(\Lambda)$, and of
meromorphic theories in general, introducing the concept of
deterministic theories.
\end{abstract}
\section{Introduction}
Recently much progress has been made
towards the classification of conformal field theories (CFT's). One approach to
this
problem is to study
the algebra of the fusion rules of representations of some chiral algebra (an
extension of the Virasoro algebra)\cite{Verlinde}. However, we are
interested here in
CFT's which would be overlooked by this technique, {\it
i.e.} theories whose fusion
rules are trivial, though they themselves are not necessarily without an
interesting structure. (Indeed, one such theory is the
natural module $V^\natural$ for the Monster,
first constructed by Frenkel, Lepowsky and
Meurman\cite{FLMbook,FLMacadsci,FLMproc}.)
Such theories must clearly be classified separately
from the mainstream if fusion rule techniques are to be employed.

We consider chiral bosonic meromorphic CFT's defined on the Riemann sphere (see
\cite{PGmer,thesis} for the relevant definitions, and also \cite{FHL}
for a mathematicians' viewpoint). We define a CFT
$\Hil$ to be self-dual if the
partition function
\begin{equation}
\chi_\Hil(\tau)={\rm Tr}_\Hil\,q^{L_0-c/24}\,,
\end{equation}
where $q=e^{2\pi i\tau}$, is covariant under modular transformations,
{\em i.e.} invariant under $S:\tau\mapsto-1/\tau$ and
invariant up to a phase under $T:\tau
\mapsto\tau+1$. (So that the full partition function when the
antichiral sector is included is then modular invariant, as required
for the theory to be physically well-defined on the torus described by
the parameter $\tau$.) This restricts us to $c\in 8\ze$. The theories
for $c=8$ and $c=16$ are easily classified\cite{PGmer}. They are
simply the FKS constructions $\Hil(\Lambda)$ from the even self-dual
lattices in the corresponding dimensions, {\em i.e.} the root lattice
of $E_8$ in 8 dimensions and the lattices ${E_8}^2$ and ${D_{16}}^+$
(an extension of the root lattice of $D_{16}$ by adding in one of the
spinor weights) in 16 dimensions.

Only in 24 dimensions does the problem of classification first become
non-trivial. Indeed, it may be argued that, since the classification
of even the even self-dual lattices in more than 24 dimensions is
intractable at present due to the rapid increase in their number with
dimension, then
$c=24$ is really the only case amenable to consideration, as well as
being of physical significance in that it is relevant to the
classification of heterotic string theories\cite{Schell:Venkov}.
There are 24
inequivalent even self-dual lattices in 24 dimensions\cite{ConSlo}.
The constructions
$\Hil(\Lambda)$ and $\widetilde\Hil(\Lambda)$ (the reflection-twisted
orbifold of $\Hil(\Lambda)$) of \cite{DGMtwisted} would thus
be naively expected to produce 48 CFT's. However, it is shown in
\cite{DGMtrialsumm}
and \cite{thesis} that the constructions produce equivalent theories if
and only if there is a corresponding doubly-even self-dual binary code. There
are
9 such codes in 24 dimensions\cite{binary:class}, and so we obtain 39 distinct
self-dual $c=24$ CFT's. These are, so far, the only such theories which have
been constructed explicitly, though in \cite{SchellYank:curious} two further
theories were postulated to exist. We shall discuss these further in a later
section.
\section{Classification of Kac-Moody algebras at $c=24$}
\label{Schell}
In \cite{Schell:Venkov} Schellekens proved results for $c=24$
self-dual CFT's analogous to those
obtained by Venkov in his reformulation of Niemeier's classification of even
self-dual lattices\cite{Venkov,ConSlo}.
In particular, it was shown that the Kac-Moody
algebra generated by the modes of the weight one states\cite{PGmer}
is restricted to contain components whose central charges sum to 24
and moreover have a common value for the ratio of the Coxeter number $g$ to the
level $k$, given in terms of the number $N$ of weight one states by
\begin{equation}
{g\over k}={N\over{24}}-1\,.
\label{hoojah}
\end{equation}
The possible combinations of algebras thus allowed in a $c=24$ self-dual
CFT with $N$
weight one states may thus be calculated. In a subsequent
paper\cite{SchellComplete}, Schellekens provided further restrictions
by an examination of higher order terms from his previous arguments,
which appear to eliminate all accidental solutions to \reg{hoojah}.
[Note that this is a departure from the analogies with the work of Venkov.]
We list the surviving possibilities below.
The rank is also indicated
for convenient reference. A $*$ indicates that the theory is one of the 39
obtained by the FKS construction or a reflection twist of such a
theory.
The two theories marked by a $\dagger$ are those claimed (but not proven)
to exist in \cite{SchellYank:curious}. Their evidence consisted of the
construction
of a modular invariant combination of characters of the corresponding Kac-Moody
algebras, though no explicit constructions were given. We shall return to this
question
in a later section.
We mark by $\oplus$ new theories proposed/constructed in this work.

\begin{tabbing}
$N\hskip30pt$  \= algebra$\hskip70pt$ \= rank\hskip50pt \=
$N\hskip30pt$  \= algebra$\hskip70pt$ \= rank \\
$0$ \> $\emptyset$ \> 0 * \>
$24$ \> $U(1)^{24}$ \> 24 * \\
$36$ \> ${C_{4,10}}$ \> 4 \>
$36$ \> ${A_{2,6}}{D_{4,12}}$ \> 6 \\
$36$ \> ${A_{1,4}}^{12}$ \> 12 * \>
$48$ \> ${A_{6,7}}$ \> 6 \\
$48$ \> ${A_{4,5}}^{2}$ \> 8 \>
$48$ \> ${A_{2,3}}^{6}$ \> 12 \\
$48$ \> ${A_{1,2}}{D_{5,8}}$ \> 6 \>
$48$ \> ${A_{1,2}}{A_{5,6}}{C_{2,3}}$ \> 8 \\
$48$ \> ${A_{1,2}}{A_{3,4}}^{3}$ \> 10 \>
$48$ \> ${A_{1,2}}^{16}$ \> 16 * \\
$60$ \> ${C_{2,2}}^{6}$ \> 12 * \>
$60$ \> ${A_{2,2}}{F_{4,6}}$ \> 6 $\dagger$ \\
$60$ \> ${A_{2,2}}^{4}{D_{4,4}}$ \> 12 \>
$72$ \> ${A_{1,1}}{C_{5,3}}{G_{2,2}}$ \> 8 \\
$72$ \> ${A_{1,1}}^{2}{D_{6,5}}$ \> 8 \>
$72$ \> ${A_{1,1}}^{2}{C_{3,2}}{D_{5,4}}$ \> 10 \\
$72$ \> ${A_{1,1}}^{3}{A_{7,4}}$ \> 10 \>
$72$ \> ${A_{1,1}}^{3}{A_{5,3}}{D_{4,3}}$ \> 12 \\
$72$ \> ${A_{1,1}}^{4}{A_{3,2}}^{4}$ \> 16 * \>
$72$ \> ${A_{1,1}}^{24}$ \> 24 * \\
$84$ \> ${B_{3,2}}^{4}$ \> 12 * \>
$84$ \> ${A_{4,2}}^{2}{C_{4,2}}$ \> 12 \\
$96$ \> ${C_{2,1}}^{4}{D_{4,2}}^{2}$ \> 16 * \>
$96$ \> ${A_{2,1}}{C_{2,1}}{E_{6,4}}$ \> 10 \\
$96$ \> ${A_{2,1}}^{2}{A_{8,3}}$ \> 12 \>
$96$ \> ${A_{2,1}}^{2}{A_{5,2}}^{2}{C_{2,1}}$ \> 16 \\
$96$ \> ${A_{2,1}}^{12}$ \> 24 * \>
$108$ \> ${B_{4,2}}^{3}$ \> 12 * \\
$120$ \> ${E_{6,3}}{G_{2,1}}^{3}$ \> 12 $\oplus$ \>
$120$ \> ${A_{3,1}}{D_{7,3}}{G_{2,1}}$ \> 12 \\
$120$ \> ${A_{3,1}}{C_{7,2}}$ \> 10 \>
$120$ \> ${A_{3,1}}{A_{7,2}}{C_{3,1}}^{2}$ \> 16 \\
$120$ \> ${A_{3,1}}^{2}{D_{5,2}}^{2}$ \> 16 * \>
$120$ \> ${A_{3,1}}^{8}$ \> 24 * \\
$132$ \> ${A_{8,2}}{F_{4,2}}$ \> 12 \>
$144$ \> ${C_{4,1}}^{4}$ \> 16 * \\
$144$ \> ${B_{3,1}}^{2}{C_{4,1}}{D_{6,2}}$ \> 16 * \>
$144$ \> ${A_{4,1}}{A_{9,2}}{B_{3,1}}$ \> 16 \\
$144$ \> ${A_{4,1}}^{6}$ \> 24 * \>
$156$ \> ${B_{6,2}}^{2}$ \> 12 * \\
$168$ \> ${D_{4,1}}^{6}$ \> 24 * \>
$168$ \> ${A_{5,1}}{E_{7,3}}$ \> 12 \\
$168$ \> ${A_{5,1}}{C_{5,1}}{E_{6,2}}$ \> 16 $\oplus$ \>
$168$ \> ${A_{5,1}}^{4}{D_{4,1}}$ \> 24 * \\
$192$ \> ${B_{4,1}}{C_{6,1}}^{2}$ \> 16 \>
$192$ \> ${B_{4,1}}^{2}{D_{8,2}}$ \> 16 * \\
$192$ \> ${A_{6,1}}^{4}$ \> 24 * \>
$216$ \> ${A_{7,1}}{D_{9,2}}$ \> 16 * \\
$216$ \> ${A_{7,1}}^{2}{D_{5,1}}^{2}$ \> 24 * \>
$240$ \> ${C_{8,1}}{F_{4,1}}^{2}$ \> 16 \\
$240$ \> ${B_{5,1}}{E_{7,2}}{F_{4,1}}$ \> 16 \>
$240$ \> ${A_{8,1}}^{3}$ \> 24 * \\
$264$ \> ${D_{6,1}}^{4}$ \> 24 * \>
$264$ \> ${A_{9,1}}^{2}{D_{6,1}}$ \> 24 * \\
$288$ \> ${B_{6,1}}{C_{10,1}}$ \> 16 \>
$300$ \> ${B_{12,2}}$ \> 12 * \\
$312$ \> ${E_{6,1}}^{4}$ \> 24 * \>
$312$ \> ${A_{11,1}}{D_{7,1}}{E_{6,1}}$ \> 24 * \\
$336$ \> ${A_{12,1}}^{2}$ \> 24 * \>
$360$ \> ${D_{8,1}}^{3}$ \> 24 * \\
$384$ \> ${B_{8,1}}{E_{8,2}}$ \> 16 $\dagger$ $\oplus$\>
$408$ \> ${A_{15,1}}{D_{9,1}}$ \> 24 * \\
$456$ \> ${D_{10,1}}{E_{7,1}}^{2}$ \> 24 * \>
$456$ \> ${A_{17,1}}{E_{7,1}}$ \> 24 * \\
$552$ \> ${D_{12,1}}^{2}$ \> 24 * \>
$624$ \> ${A_{24,1}}$ \> 24 * \\
$744$ \> ${E_{8,1}}^{3}$ \> 24 * \>
$744$ \> ${D_{16,1}}{E_{8,1}}$ \> 24 * \\
$1128$ \> ${D_{24,1}}$ \> 24
\end{tabbing}
\section{Complementary representations}
Consider a bosonic meromorphic hermitian CFT $S$ which
may be extended to form a self-dual theory by adding in a representation $U$
(real, hermitian and satisfying the additional locality requirement as detailed
in \cite{thesis,DGMtriality}), {\em i.e.}
${\cal H}=S\oplus U$ is self-dual.
\subsection{Definition}
Now, let $\theta$ be the automorphism of ${\cal H}$ defined to be $1$ on $S$
and
$-1$ on $U$. The invariant sub-theory is simply $S$, and we shall assume that
we may construct a corresponding self-dual orbifold theory
${\cal H}_\theta=S\oplus
U'$, with $U'$ a representation of $S$.

We have
\begin{equation}
\chi_{{\cal H}_\theta}(\tau)=\chi_S(\tau)+\chi_{U'}(\tau)\,,
\end{equation}
where
\begin{eqnarray}
\chi_{U'}(\tau)&=&{1\over 2}\left(
{1\lower 6pt \hbox{$
{\square\atop\displaystyle\theta}$}}(\tau)+
{\theta\lower 6pt \hbox{$
{\square\atop\displaystyle\theta}$}}(\tau)\right) \nonumber \\
\chi_S(\tau)&=&{1\over 2}\left(\chi_\Hil(\tau)+{\theta
\lower 6pt \hbox{$
{\square\atop{\displaystyle 1}}$}}(\tau)\right)\\
\end{eqnarray}
and all the boxes are to be understood with reference to $\Hil$.

Thus
\begin{equation}
\chi_S(S(\tau))={1\over 2}\left(\chi_\Hil(\tau)+{1
\lower 6pt \hbox{$
{\square\atop\displaystyle\theta}$}}(\tau)\right)
\end{equation}
and
\begin{equation}
\chi_S(ST(\tau))={1\over 2}\left(\chi_\Hil(\tau)+{\theta
\lower 6pt \hbox{$
{\square\atop\displaystyle\theta}$}}(\tau)\right)\,.
\end{equation}
Hence
\begin{equation}
\chi_\Hil(\tau)+\chi_{{\cal H}_\theta}(\tau)=\chi_S(\tau)+\chi_S(S(\tau))+
\chi_S(ST(\tau))\,.
\label{sum}
\end{equation}
In other words, the sum of the partition functions of $\Hil$ and
$\Hil_\theta$ is determined solely
in terms of the partition function of the invariant sub-theory $S$.
(For $c=24$, the partition functions are restricted to be of the form
$J(\tau)+N$,
where $J$ is the elliptic modular function with zero constant term and $N$ is
the
number of weight one states.)

\proclaim Definition. We say that $U'$ is the complementary representation
of $S$ to $U$. Further, if $U$ and $U'$ are equivalent as representations of
$S$,
we say that $U$ is self-complementary with respect to $S$. \par

\subsection{Example}
Consider $S=\Hil(\Lambda)_+$, $\Lambda$ an even self-dual
lattice. ($\Hil(\Lambda)_+$ is the sub-theory invariant
under the involution on $\Hil(\Lambda)$ given by lifting the
reflection symmetry of the lattice.)
We have shown in \cite{DGMtriality} (though this paper has yet to be
published, the result in the specific case in which $\Lambda$ is the
Leech lattice has been proven by Dong \cite{Dongmoonrep})
that there are exactly two representations
$U_1$ and $U_2$ extending $S$ to a self-dual CFT,
{\em i.e.} $\Hil(\Lambda)\cong S\oplus U_1$ and $\widetilde\Hil(\Lambda)
\cong S\oplus U_2$.
We can argue that ${U_1}'\cong U_2$. Suppose not. Then $U_1$ must be
self-complementary, and so the number of weight one states in $U_1$
would then be fixed by the partition function of $S$. Now, assuming that the
notion of complementarity is symmetric,{\em i.e.} $U''\cong U$
(see below for more discussion),
$U_2$ must also be self-complementary (otherwise ${U_2}'\cong U_1$, and hence
${U_1}'\cong{U_2}''\cong U_2$). So \reg{sum} then implies that $U_2$ must
have the same number of weight one states as $U_1$. We know from the
explicit constructions of these representations that this is a contradiction,
as required.

Note the advantage of this point of view is that it puts $\Hil(\Lambda)$
and $\widetilde\Hil(\Lambda)$ on an equal footing, unlike the conventional
approach.
\subsection{Symmetry of the definition}
We discuss here the property alluded to in the above example.
Suppose $U'$ is the complementary representation of the CFT $S$ to a
representation $U$. We can then construct a representation $U''$ of S
complementary to $U'$. We see from \reg{sum} that it must have the
same partition function as $U$ (which, for $c>24$, is more than saying merely
that the number of weight one states is the same).
This observation alone is enough in the case of our example above,
since we know that $U_1$ and $U_2$ have distinct partition functions and
so symmetry is assured. The question in general of whether $U\cong U''$
remains an open one however. Even for $c=24$, we have examples of
distinct theories with the same number of weight one states, and so
the equality of the partition functions alone is not sufficient.
\subsection{Application to self-dual $c=24$ conformal field theories}
\label{applic} We may ask exactly what we must know of $\chi_S(\tau)$
in order to calculate the sum in \reg{sum}. By
itself, \reg{sum} is of limited use for evaluating the partition function of
the orbifold theory as typically we have insufficient knowledge of the
transformation properties of
$\chi_s(\tau)$.
However, for central charge 24, the situation is
considerably simplified. In particular,
we need only consider the number of states
of conformal weight one.

Clearly, $\chi_S(\tau)$ is a $\Gamma_0(2)$ invariant, {\em i.e.} it is
invariant under both $T$ and $ST^2S$. So we can write\cite{Tuite:moon}
\begin{equation}
\chi_S(\tau)=N_S+{1\over 2}\left\{J(\tau)+\alpha\left(
{\theta_3(\tau)^8\theta_4(\tau)^8+2^{-4}\theta_2(\tau)^{16}\over
{\eta(\tau)^8\eta(2\tau)^8}}+24\right)
+\beta\left(\left({\eta(\tau)\over{\eta(2\tau)}}\right)^{24}+24\right)\right\}\,,
\end{equation}
using the $\Gamma_0(2)$ and $\Gamma_0(2)+$ hauptmoduls,
where $N_S$ is the number of states of weight 1 in $S$ and $\alpha+\beta=1$
(since $\chi_S(\tau)\sim q^{-1}$ as $q\rightarrow 0$).
(Note that if $\alpha=0$ or $1$ then
the Moonshine
Conjecture\cite{moonshine} holds in this case, {\em i.e.}
${\theta
\lower 6pt \hbox{$
{\square\atop{\displaystyle 1}}$}}(\tau)$
is a
hauptmodul for a genus zero modular group.)
Considering the transformation $S:\tau\mapsto-1/\tau$ shows that
$\alpha$ is to be interpreted as the number of weight ${1\over 2}$ states in
the
twisted sector before projection onto a meromorphic theory, and is given by
\begin{equation}
\alpha={N_{S,2}-98580\over{2048}}\,,
\label{alpha}
\end{equation}
where $N_{S,2}$ is the number of states of weight 2 in $S$.
We also find, using \reg{sum}, that
\begin{equation}
N_\Hil+N_{\Hil'}=3N_S+24(1-\alpha)\,,
\label{beta}
\end{equation}
where $N_\Hil$ is the number of weight one states in $\Hil=S\oplus
U$, and similarly for $N_{\Hil'}$.

In a section 5, we will apply our arguments to cases in which
the theory $S$ is constructed as
the invariant sub-theory of a
self-dual theory under an involution, and we are thus simply
considering the
orbifold of the first theory with respect to this involution.
The representations of the sub-theory giving us the orbifold and
the original theory are complementary, allowing us to use
\reg{alpha} and \reg{beta} to calculate the number of weight one states in the
orbifold theory.
Together with a knowledge of the Kac-Moody algebra corresponding to
the weight one states of $S$, the number
of weight one states allows us to look up in the table given by the
results of Schellekens in section \ref{Schell} and identify the most
likely candidate for the theory.
An absence of a suitable candidate will imply that the orbifold cannot be
consistent.

[The motivation for the above approach is based upon an extension of
the analogies between constructions of lattices from binary codes and
constructions of CFT's from lattices which were summarised in
\cite{DGMtrialsumm}. In \cite{PSMcodes} constructions for all of the
Niemeier lattices from ternary codes were given, suggesting the
existence of corresponding constructions of CFT's from (Eisenstein)
lattices by some form of $\ze_2$-orbifold approach. The exact nature
of the orbifold theory would be difficult to write down. Instead it is
considerably easier to postulate some sub-theory with $c=24$ of the
$c=48$ theory
$\Hil(\Lambda)$ ($\Lambda$ the 48-dimensional even self-dual lattice
corresponding to a 24-dimensional Eisenstein lattice) which is to form
the invariant space upon which the
orbifold is constructed. The above argument would then give us the sum
of the number of weight one states in two (potentially distinct)
orbifolds that may be then formed by the addition of complementary
representations, and reference to the Kac-Moody
algebra of the invariant theory and Schellekens' results would
potentially provide
information sufficient to identify the theories. This is a program
which is still in progress.]
\section{Generalisation to automorphisms of order $>2$}
\label{qwer}
Let $\theta$ be an automorphism of order $N$ on a
CFT $\Hil$.
Then, assuming that the orbifold $\Hil_\theta$ of $\Hil$ with respect
to
$\theta$ is consistent,
its partition function is given by
\begin{eqnarray}
\chi_{\Hil_\theta}(\tau)&=&{1\over N}\left(
{1\lower 6pt \hbox{$
{\square\atop\displaystyle 1}$}}(\tau)+
{\theta\lower 6pt \hbox{$
{\square\atop\displaystyle 1}$}}(\tau)+
\cdots +
{\theta^{N-1}\lower 6pt \hbox{$
{\square\atop\displaystyle 1}$}}(\tau)\right.\nonumber\\
&&+{1\lower 6pt \hbox{$
{\square\atop\displaystyle\theta}$}}(\tau)+
\cdots
+{\theta^{N-1}\lower 6pt \hbox{$
{\square\atop\displaystyle\theta}$}}(\tau)\nonumber\\
&&+\cdots\nonumber\\
&&\left.+{1\lower 6pt \hbox{$
{\square\atop\displaystyle\theta^{N-1}}$}}(\tau)+
\cdots
+{\theta^{N-1}\lower 6pt \hbox{$
{\square\atop\displaystyle\theta^{N-1}}$}}(\tau)
\right)\,.
\end{eqnarray}
[Note that we are assuming the absence of a global phase anomaly as
described in \cite{Tuite:moon,Vafa}, {\em i.e.} we are assuming that the
conformal weights in the twisted sectors lie in $\ze/N$.]
Set
\begin{equation}
S(\tau)={1\over N}\left(
{1\lower 6pt \hbox{$
{\square\atop\displaystyle 1}$}}(\tau)+
{\theta\lower 6pt \hbox{$
{\square\atop\displaystyle 1}$}}(\tau)+
\cdots +
{\theta^{N-1}\lower 6pt \hbox{$
{\square\atop\displaystyle 1}$}}(\tau)\right)\,.
\label{gibbon}
\end{equation}
Then
\begin{eqnarray}
&S(\tau)+S(S\tau)+S(TS\tau)+\ldots+S(T^{N-1}S\tau)=\nonumber\\
&{1\over N}\left({1\lower 6pt \hbox{$
{\square\atop\displaystyle 1}$}}(\tau)+
{\theta\lower 6pt \hbox{$
{\square\atop\displaystyle 1}$}}(\tau)+
\cdots +
{\theta^{N-1}\lower 6pt \hbox{$
{\square\atop\displaystyle 1}$}}(\tau)
+{1\lower 6pt \hbox{$
{\square\atop\displaystyle 1}$}}(\tau)+
{1\lower 6pt \hbox{$
{\square\atop\displaystyle\theta}$}}(\tau)+
\cdots +
{1\lower 6pt \hbox{$
{\square\atop\displaystyle\theta^{N-1}}$}}(\tau)+\right.\nonumber\\
&\left.{1\lower 6pt \hbox{$
{\square\atop\displaystyle 1}$}}(\tau)+
{\theta^{N-1}\lower 6pt \hbox{$
{\square\atop\displaystyle\theta}$}}(\tau)+
\cdots +
{\theta\lower 6pt \hbox{$
{\square\atop\displaystyle\theta^{N-1}}$}}(\tau)
+\cdots+{1\lower 6pt \hbox{$
{\square\atop\displaystyle 1}$}}(\tau)+
{\theta\lower 6pt \hbox{$
{\square\atop\displaystyle\theta}$}}(\tau)+
\cdots +
{\theta^{N-1}\lower 6pt \hbox{$
{\square\atop\displaystyle\theta^{N-1}}$}}(\tau)\right)\,.
\end{eqnarray}
We see therefore that, for $N$ prime (say$N=p$),
\begin{equation}
\chi_{\Hil_\theta}(\tau)+\chi_\Hil(\tau)=
S(\tau)+S(S\tau)+S(TS\tau)+\ldots+S(T^{p-1}S\tau)\,.
\label{result}
\end{equation}
(One can trivially see that
the restriction to $N$ prime is not simply a device to render the
calculation tractable.)

Let us restrict ourselves from now on to the case of a $c=24$ CFT
with $(p-1)|24$. For such $p$, $\Gamma_0(p)$
is a genus zero modular group. Restricting further to $p$ a divisor of
the order of the Monster group gives
us that $\Gamma_0(p)+$ (the extension of $\Gamma_0(p)$ by the Atkin-Lehner
involution $S_p:\tau\mapsto -1/p\tau$)
is also of genus zero, and if we take
$(p+1)|24$ then we have a convenient  explicit form for the
corresponding hauptmodul.
Since we have considered the
case $p=2$ already in the previous section, then we simply consider
$p=3$, 5 or 7.
Define $r$ and $s$ by
$(p-1)r=24$, $(p+1)s=24$. Then the relevant hauptmoduls can be taken
to be given by\cite{Tuite:moon}
\begin{eqnarray}
f_p(\tau)&=&\left({\eta(\tau)\over{\eta(p\tau)}}\right)^r+r\nonumber\\
g_p(\tau)&=&\psi_p(\tau)+{2^s\over{\psi_p(\tau/2)}}
-2^s{\psi_p(\tau/2)\over{\psi_p(\tau)}}+s
\end{eqnarray}
with
\begin{equation}
\psi_p(\tau)=\left({\eta(\tau)\eta(p\tau)\over{\eta(2\tau)\eta(2p\tau)}}
\right)^s\,.
\end{equation}
We know that (still assuming that no phase anomaly is present)
$S(\tau)$ is $\Gamma_0(p)$ invariant. It can thus be written as a
rational function of $f_p(\tau)$. However, let us instead try to write
it in the following form:
\begin{equation}
S(\tau)={1\over p}J(\tau)+\sum_{i=1}^{p-1}\alpha_if_p(\tau)^i+
\sum_{j=1}^{p-1}\beta_jg_p(\tau)^j+C\,,
\label{thingy}
\end{equation}
for some constant $C$.
[Note that in the case considered by Tuite in \cite{Tuite:moon},
Monstrous Moonshine implies that $\alpha_1={p-1\over p}$ or
$\beta_1={p-1\over p}$ and that the remaining coefficients vanish. The
situation for a CFT other than $V^\natural$ as
studied here is, as we see, more intricate.]
$S(\tau)$ only has singularities at the parabolic cusps of ${\cal
F}_p=H/\Gamma_0(p)$ ($H$ the upper half plane). We make the particular
choice of inequivalent cusps as $\{0,i\infty\}$. These are mapped onto
each other by $S_p$.
If we can choose $\{\alpha_i\}$ and $\{\beta_j\}$ to match the
singularities of $S(\tau)$ with those of the right hand side of
\reg{thingy}, then, by Liouville's theorem, $S(\tau)$ must be given by
\reg{thingy} up to an additive constant.
Now $S(\tau)\sim q^{-1}$ as $q\rightarrow 0$, and
\begin{eqnarray}
S(-1/p\tau)&=&{1\over p}\left({1\lower 6pt \hbox{$
{\square\atop\displaystyle 1}$}}(p\tau)+
{1\lower 6pt \hbox{$
{\square\atop\displaystyle\theta}$}}(p\tau)+\cdots
+{1\lower 6pt \hbox{$
{\square\atop\displaystyle\theta^{p-1}}$}}(p\tau)\right)\nonumber\\
&\widetilde{q\rightarrow 0}&{1\over p}q^{-p}+M_1q^{-(p-1)}+M_2q^{-(p-2)}+\ldots
+M_{p-1}q^{-1}\,,
\end{eqnarray}
where $M_j={1\over p}\sum_{i=1}^{p-1}M_{j,i}$ with $M_{j,i}$ the
number of states of conformal weight $j/p$ in the sector twisted by $\theta^i$.
Also, we have $f_p(\tau)$, $g_p(\tau)\sim q^{-1}$ and
$f_p(-1/p\tau)\sim q$, $g_p(-1/p\tau)=g_p(\tau)\sim q^{-1}$ as $q\rightarrow
0$.\,
while $J(\tau)\sim q^{-1}$, $J(-1/p\tau)\sim q^{-p}+O(1)$.

We attempt to match the singularities. {\em A priori} there is no reason to
suppose that this is possible, though we find that it is, and in fact for $p=3$
\begin{eqnarray}
\alpha_1=2/3-M_2\,;\qquad\alpha_2=-M_1\nonumber\\
\beta_1=M_2\,;\qquad\beta_2=M_1\,,
\end{eqnarray}
with similar results for $p=5$, 7.

Now that we know that it is possible to express $S(\tau)$ in the form
\reg{thingy},
let us evaluate $\{\alpha_i\}$ and $\{\beta_j\}$ in terms of the
coefficients
$P_i$ in the series expansion
of $S(\tau)$ in terms of $q$,
{\em i.e.} taking
\begin{equation}
S(\tau)=\sum_{n=0}^\infty P_nq^{n-1}\,.
\end{equation}
We may also evaluate the constant $C$ in \reg{thingy} in terms of the $P_i$.

Putting together these results, we may evaluate the constant term on
the right hand side of \reg{result}, which we know to be of the form
$J(\tau)+{\rm constant}$. From \reg{thingy} and the known
transformation properties of the functions $J$, $f_p$ and $g_p$, we have that
\begin{equation}
S(S\tau)={1\over
p}J(p\tau)+\sum_{i=1}^{p-1}\alpha_i\left[\left(p^{1\over
2}{\eta(\tau)\over
{\eta\left({\tau\over
p}\right)}}\right)^r+r\right]^i+\sum_{j=1}^{p-1}\beta_jg_p\left({\tau\over
p}\right)^j+C\,.
\end{equation}
The sum $S(S\tau)+\cdots+S(T^{p-1}S\tau)$ simply projects out the
meromorphic piece of $S(S\tau)$, {\em i.e.} it just gives us $p$
times the constant term in $S(S\tau)$. Thus, the constant term on the
right hand side of \reg{result} is
\begin{equation}
C_0=N_1+p\left(\sum_{i=1}^{p-1}\alpha_ir^i+\sum_{j=1}^{p-1}\beta_jC\left[g_p(\tau)^j\right]+C\right)\,,
\end{equation}
where $C[\cdot]$ denotes the constant term in the series expansion in
terms of $q$
[we note that $\left(\eta(\tau)/\eta(\tau/p)\right)^r$ is of $O(q^{1/p})$ as
$q\rightarrow 0$].
For $p=3$ this becomes
\begin{eqnarray}
C_0=4\left({19683P_1-327P_2+7P_3-28561526\over{19683}}\right)\,.
\label{star}
\end{eqnarray}
Similarly, for $p=5$ we get
\begin{eqnarray}
C_0&=&6\left(9765625P_1-714248P_2+73911P_3-5023P_4+\right.\nonumber\\
&&\left.161P_5-73197587534\right)/9765625\,,
\label{mac}
\end{eqnarray}
and for $p=7$
\begin{eqnarray}
C_0&=&4\left(13040259233946+564950498P_1-64391796P_2+7019075P_3-\right.\nonumber\\
&&\left.112200P_4-91497P_5+12793P_6-598P_7\right)/282475249\,.
\label{mac2}
\end{eqnarray}
\reg{star}, \reg{mac} and \reg{mac2} constitute the main concrete
result of this paper, though it is in some sense the ideas behind the
techniques which are of import. In any case, they support the validity
of our conjectured technique.
\section{Applications}
In this section, we shall provide a few examples of the application of
the above results to determine the orbifolds obtained from known
CFT's at $c=24$. The only self-dual theories which
have so far been shown to exist at this value of the central charge
are the theories $\Hil(\Lambda)$ and
$\widetilde\Hil(\Lambda)$\cite{DGMtwisted} for $\Lambda$ an even
self-dual lattice.
Note that this section makes no pretense at completeness, and indeed
the examples given in sections 5.2 and 5.3 are trivial,  but merely
gives a flavour of the ongoing program, as well as demonstrating in
the first subsection the potential importance of this approach
regarding the completion of the full classification problem.
\subsection{Automorphism group of the theories $\Hil(\Lambda)$ and
$\widetilde\Hil(\Lambda)$}
\label{flobtip}
Firstly, we must identify the automorphisms of $\Hil(\Lambda)$ and
$\widetilde\Hil(\Lambda)$. This is a problem which has so far remained
surprisingly unsolved, although we may write down an obvious subgroup
of the automorphism group of each induced by automorphisms of the
underlying lattice.

Let
\begin{equation}
\hat C(\Lambda)=\{(R,S):R\in\Aut(\Lambda),\,S\gamma_\lambda S^{-1}=v_{R,S}
(\lambda)\gamma_{R\lambda}\}\,,
\end{equation}
for some $v_{R,S}(\lambda)=\pm 1$ and $\gamma_\lambda\gamma_\mu=(-1)^{
\lambda\cdot\mu}\gamma_\mu\gamma_\lambda$, $\lambda$, $\mu\in\Lambda$.
The kernel of the homomorphism $(R,S)\mapsto R$ is
$\Gamma(\Lambda)=\{\pm\gamma_\lambda:\lambda\in\Lambda\}$,
and so we have the exact sequence
\begin{equation}
1\rightarrow\Gamma(\Lambda)\rightarrow\hat C(\Lambda)\rightarrow
\Aut(\Lambda)\rightarrow 1\,.
\end{equation}
This provides a group of automorphisms of $\Hil(\Lambda)$ and
$\widetilde\Hil(\Lambda)$
given by
\begin{eqnarray}
\label{amv}
u_{R,S}a_n^ju_{R,S}^{-1}=R_{ij}a_n^i\,,&u_{R,S}c_r^ju_{R,S}^{-1}=
R_{ij}c_r^i\,,\nonumber\\
u_{R,S}|\lambda\rangle=v_{R,S}(\lambda)|R\lambda\rangle\,,&
u_{R,S}\chi=S\chi\,.
\end{eqnarray}
(See \cite{DGMtwisted} for the notation.)
Now
$\iota_P=(-1,1)\in
\hat C(\Lambda)$ acts trivially on $\widetilde\Hil(\Lambda)$. Thus, we have a
group
of automorphisms $C(\Lambda)\cong\hat C(\Lambda)/\hat\ze_P$, where
$\hat\ze_P\equiv\{(\pm 1,1)\}\cong\ze_2$, and a homomorphism $(\pm R,S)\mapsto
R$ of $C(\Lambda)\rightarrow\Aut(\Lambda)/\ze_P$ with kernel $\Gamma(\Lambda)$,
giving us the exact sequence
\begin{equation}
1\rightarrow\Gamma(\Lambda)\rightarrow C(\Lambda)\rightarrow
\Aut(\Lambda)/\ze_P\rightarrow 1\,.
\end{equation}

The group $\Gamma(\Lambda)$ is an extra-special 2-group, {\it i.e.}
$|\Gamma(\Lambda)|=2^{d+1}$ with centre $\ze_2$ and $\Gamma(\Lambda)/\ze_2=
\Lambda/2\Lambda\cong{\ze_2}^d$.

On the other hand, it is $\iota=(1,-1)$ which has trivial action when $\hat
C(\Lambda)$ acts on $\Hil(\Lambda)$. So we have a representation of
$Co(\Lambda)=
\hat C(\Lambda)/\ze_\iota$, where $\ze_\iota=\{1,\iota\}\cong\ze_2$. The
homomorphism $(R,\pm S)\mapsto R$ leads to the exact sequence
\begin{equation}
1\rightarrow\Lambda/2\Lambda\rightarrow Co(\Lambda)\rightarrow\Aut(\Lambda)
\rightarrow 1\,.
\end{equation}
Thus, we need to consider the automorphism groups of the Niemeier
lattices. This is described in \cite{ConSlo}.

However, this is clearly not the full automorphism group. For example,
in \cite{thesis} a ``triality'' operator is constructed on the
theories $\Hil(\Lambda_\C)$ and $\widetilde\Hil(\widetilde\Lambda_\C)$
(see \cite{thesis} for the notation-the lattices are a subset of
all the Niemeier lattices) which does not map the states
$a_{-1}^i|0\rangle$ into themselves, unlike the above group of
symmetries. Further, we have\cite{thesis}
$\Hil(\widetilde\Lambda_\C)=\widetilde\Hil(\Lambda_\C)$, and the
lattice induced automorphisms of one theory clearly are not
necessarily lattice induced in the other, since the space of states
$\{a_{-1}^i|0\rangle\}$ is in general not invariant.
Indeed, it is easily shown that any automorphism of
$\Hil(\Lambda)$ which {\em does} map the states $a_{-1}^i|0\rangle$
into themselves is of the above form.
\subsubsection{Automorphisms of $\Hil(\Lambda)$}
If we write
\begin{equation}
\theta a_{-1}^i|0\rangle={R^i}_ja_{-1}^j|0\rangle\,,
\end{equation}
then
\begin{equation}
\theta a_n^i\theta^{-1}=\theta
V(a_{-1}^i|0\rangle)_n\theta^{-1}
=V(\theta a_{-1}^i|0\rangle)_n
={R^i}_ja^j_n\,.
\end{equation}
So
\begin{equation}
\lambda^i\theta|\lambda\rangle=\theta
a_0^i|\lambda\rangle
={R^i}_ja^j_0\theta|\lambda\rangle\,.
\end{equation}
But the unitarity of $\theta$ clearly implies that the matrix $R$ is
unitary. Thus
$a^i_0\theta|\lambda\rangle={R^{\dagger i}}_j\lambda^j\theta|
\lambda\rangle$. Hence $\theta|\lambda\rangle$ is a momentum
eigenstate of momentum $R^\dagger\lambda$. Now, since $\theta$
preserves conformal weights\cite{thesis} and the conformal weight of
$|\lambda\rangle$ is ${1\over 2}\lambda^2={1\over
2}(R^\dagger\lambda)^2$, we see that
$\theta|\lambda\rangle=v(\lambda)|R^\dagger\lambda\rangle$, where
$v(\lambda)$ is a scalar. But the theory is hermitian, and so we see
trivially from the definition of the hermitian structure in
\cite{thesis} that $\overline{\theta\psi}=\theta\overline\psi$.
Hence, putting $\psi=a_{-1}^i|0\rangle$, we obtain that $R$ is real.
Thus, $R\in{\rm Aut}\,(\Lambda)$ as required, and we see that $\theta$
is an automorphism of the above form.

Now consider an arbitrary finite order automorphism $\theta$ of
$\Hil(\Lambda)$. This clearly induces an automorphism $\sigma$ (of
finite order) of the Lie algebra $g$ generated by the zero modes of
the states of conformal weight one. From Proposition 8.1 of
\cite{Kac}, we know that $\sigma$ is conjugate to an automorphism
which leaves the Cartan subalgebra invariant. At this point we
must be careful, for it is not entirely clear that an arbitrary
automorphism of $g$ can be lifted to one of $\Hil(\Lambda)$ (in fact
it is not true-see chapter 7 of \cite{thesis} for an example on the
failure of an automorphism of the root lattice of $g$ to lift to an
automorphism of the corresponding Niemeier lattice). Instead,
one can say that there exists a choice of Cartan subalgebra ({\em i.e.}
not necessarily the standard choice $a_0^i$) which is invariant under
$\sigma$, with an orthonormal basis $V(\psi^i)_0$ say, $1\leq i\leq
d$. Then we can define $A^i_n=V(\psi^i)_n$, and these operators will
obey the usual Heisenberg commutation relations. By the result of
appendix G of \cite{thesis}, the CFT is of the form
$\Hil(\Lambda')$, built up using the oscillators $A^i_n$. Clearly,
$\Lambda\cong\Lambda'$. In this picture, $\theta$ leaves invariant the
Cartan subalgebra $\langle\{A^i_0\}\rangle$, and so, by the above
argument, must be of the form of a lattice induced automorphism.
Hence, in the original picture $\Hil(\Lambda)$, the automorphism
$\theta$ is conjugate to a lattice induced automorphism (the conjugacy
being by the isomorphism $\Hil(\Lambda)\cong\Hil(\Lambda')$), and so
the orbifold theory $\Hil(\Lambda)_\theta$ will be isomorphic to one
of the lattice automorphism induced orbifolds, since the construction
of an orbifold simply depends on the conjugacy class of the
automorphism\cite{Hollthesis}.
To conclude, we may
identify all
orbifolds of
$\Hil(\Lambda)$ by
simply considering
the lattice induced automorphisms.

As a simple example, let us consider the triality automorphism of the
theories $\Hil(\Lambda_\C)$ described in chapter 5 of \cite{thesis}.
These theories contain the weight one states
\begin{eqnarray}
\epsilon^i&=&a^i_{-1}|0\rangle\nonumber\\
\zeta^i&=&{i\over\sqrt 2}\left(|\sqrt 2e_i\rangle+|-\sqrt
2e_i\rangle\right)\nonumber\\
\eta^i&=&{1\over\sqrt 2}\left(|\sqrt 2e_i\rangle-|-\sqrt
2e_i\rangle\right)\,,
\end{eqnarray}
where $\{e_i\}$ are a set of orthonormal vectors. The triality
automorphism\cite{thesis} is an involution $\theta$ which
acts on these states as
\begin{equation}
\theta\epsilon^i=\zeta^i\,;\qquad
\theta\zeta^i=\epsilon^i\,;\qquad
\theta\eta^i=-\eta^i\,.
\end{equation}
Choosing $\{V(\eta^i)_0\}$ as the generators of our Cartan
subalgebra, we find
\begin{equation}
[V(\eta^i)_0,V(\rho^j_\pm)_0]=\pm\sqrt 2\delta^{ij}V(\rho^j_\pm)_0
\end{equation}
(no summation), where $\rho^i_\pm=\zeta^i\pm\eta^i$. We then see that
$\theta$ acts on these states so as to flip the sign of the momenta.
Since the dot products of lattice vectors must be preserved by the
action of $\theta$, this is sufficient to establish that $\theta$ acts
on the lattice simply by reflection in this picture. Hence, we have
established explicitly the conjecture, made
on the basis
of an analysis of the partition function of the orbifold theory
in \cite{thesis}, that $\widetilde
{\Hil(\Lambda_\C)}_\theta\cong\widetilde\Hil(\Lambda_\C)$.
\subsubsection{Automorphisms of $\widetilde\Hil(\Lambda)$}
Similarly, let us consider an automorphism $\theta$ of
$\widetilde\Hil(\Lambda)$ which maps the space spanned by the set of
states $\{a_{-1}^ia_{-1}^j|0\rangle\}$ into itself. Write
\begin{equation}
\theta
a_{-1}^ia_{-1}^j|0\rangle={S^{ij}}_{kl}a_{-1}^ka_{-1}^l|0\rangle\,.
\label{eps11}
\end{equation}
Then
\begin{eqnarray}
\lambda^i\lambda^j\theta(|\lambda\rangle+|-\lambda\rangle)&=&
\theta
V_0(a_{-1}^ia_{-1}^j|0\rangle)(|\lambda\rangle+|-\lambda\rangle)
\nonumber\\
&=&{S^{ij}}_{kl}V_0(a_{-1}^ka_{-1}^l|0\rangle)\theta
(|\lambda\rangle+|-\lambda\rangle)\,.
\end{eqnarray}
Now, unitarity and the hermitian structure give, as above, that
${S^{ij}}_{kl}$ is real and
\begin{equation}
V_0(a_{-1}^ia_{-1}^j|0\rangle)\theta(|\lambda\rangle+|-\lambda\rangle)=
{S^{ij}}_{kl}\lambda^k\lambda^l\theta(|\lambda\rangle+|-\lambda\rangle)\,,
\end{equation}
{\em i.e.} in particular,
$\theta(|\lambda\rangle+|-\lambda\rangle)$ is an eigenstate
of $V_0(a_{-1}^ia_{-1}^j|0\rangle)$ for all $i$, $j$. Consider a
24-dimensional theory with $\lambda$ of length squared two. Then the
only states of conformal weight one are of the form
$|\mu\rangle+|-\mu\rangle$, $\mu\in\Lambda(2)$.
Thus, we must have
\begin{equation}
\theta(|\lambda\rangle+|-\lambda\rangle)=\sum_{\mu\in\Lambda(2)}C_\mu
(|\mu\rangle+|-\mu\rangle)\,,
\end{equation}
and since $|\mu\rangle+|-\mu\rangle$ is an eigenstate of
$V_0(a_{-1}^ia_{-1}^j|0\rangle)$ with eigenvalue $\mu^i\mu^j$, and
further $\mu^i\mu^j={S^{ij}}_{kl}\lambda^k\lambda^l$ (together with
the known value of $\mu^2$) determines $\mu$
(up to a sign)(since we may write the ratios $\mu^j/\mu^1$ in
terms of $S$ and $\lambda$), we clearly have
\begin{equation}
\theta(|\lambda\rangle+|-\lambda\rangle)=C_\mu
(|\mu\rangle+|-\mu\rangle)\,,
\end{equation}
for the unique (up to sign) $\mu\in\Lambda(2)$ such that
$\mu^i\mu^j={S^{ij}}_{kl}\lambda^k\lambda^l$ and $\mu^2=\lambda^2$.
For $\lambda$ of length squared 4, we may also have states of the form
$a_{-1}^i(|\mu\rangle-|-\mu\rangle)$, $a_{-1}^ia_{-1}^j|0\rangle$ and
$c_{-1/2}^i\chi$. These three separate sets of states are mapped into
themselves under the action of $V_0(a_{-1}^ia_{-1}^j|0\rangle)$. So,
any component of $\theta(|\lambda\rangle+|-\lambda\rangle)$ in these
spaces must itself be an eigenvector of
$V_0(a_{-1}^ia_{-1}^j|0\rangle)$, for all $i$, $j$.
However
\begin{equation}
V_0(a_{-1}^ia_{-1}^j|0\rangle)\sum_kc_{-1/2}^k\chi_k=
{1\over 2}\left(c_{-1/2}^j\chi_i+c_{-1/2}^i\chi_j\right).
\end{equation}
Thus, we require that the twisted sector ground states $\chi_k$ vanish
for $k\neq i$, $j$ and that $\chi_i=\chi_j$. This must be true for all
$i$ and $j$. Hence we trivially see that there can be no component of
$\theta(|\lambda\rangle+|-\lambda\rangle)$ lying in the twisted sector
for $\lambda$ of length squared 4.
A similar argument holds in general, and we see that
\begin{equation}
\theta(|\lambda\rangle+|-\lambda\rangle)=
\sum_\mu C_\mu(|\mu\rangle+|-\mu\rangle)\,.
\end{equation}
Then, as argued above,
\begin{equation}
\theta(|\lambda\rangle+|-\lambda\rangle)=
C_\lambda(|\mu(\lambda)\rangle+|-\mu(\lambda)\rangle)\,,
\end{equation}
where $\mu(\lambda)$ is the unique (up to a sign) (lattice) vector
such that
\begin{equation}
\mu^i(\lambda)\mu^j(\lambda)={S^{ij}}_{kl}\lambda^k\lambda^l
\label{del11}
\end{equation}
and
$\mu(\lambda)^2=\lambda^2$.

Considering the action of $\theta$ on
$V(|\lambda\rangle+|-\lambda\rangle,z)
(|\rho\rangle+|-\rho\rangle)$ for $\lambda$, $\rho\in\Lambda$ clearly
gives us
\begin{equation}
\mu(\lambda+\rho)=\pm(\mu(\lambda)\pm\mu(\rho))\,.
\label{del12}
\end{equation}
If we can then make a consistent choice of signs for $\mu(\lambda)$,
$\lambda\in\Lambda$, such that
\begin{equation}
\mu(\lambda+\rho)=\mu(\lambda)+\mu(\rho)
\label{frtgyh}
\end{equation}
for all $\lambda$, $\rho\in\Lambda$ (see appendix \ref{app11}),
we will have that $\mu(\lambda)$
is linear in $\lambda$ and hence of the form
\begin{equation}
\mu^i(\lambda)={R^i}_j\lambda^j
\label{gam11}
\end{equation}
for some (constant) matrix $R$. Since $\mu(\lambda)\in\Lambda$ for all
$\lambda\in\Lambda$ and $\mu(\lambda)^2=\lambda^2$, $R\in{\rm Aut}(\Lambda)$.

Now, the action of $\theta$ on the whole CFT is
determined by its action on the states $a_{-1}^ia_{-1}^j|0\rangle$,
$|\lambda\rangle+|-\lambda\rangle$ (for $\lambda$ in a set of basis
vectors for $\Lambda$), which generate $\Hil(\Lambda)_+$,
and a single state in the twisted sector (which
forms an irreducible module of $\Hil(\Lambda)_+$), since the modes of
the vertex
operators corresponding to these states generate the CFT\cite{thesis}.
So, in order to verify that $\theta$ is lattice induced, we need
merely check the form of its action upon those states.

$\theta\overline\psi=\overline{\theta\psi}$ gives us, taking
$\psi=|\lambda\rangle+|-\lambda\rangle$, ${C_\lambda}^\ast=C_\lambda$.
Also unitarity of $\theta$ requires $|C_\lambda|=1$, and so
$C_\lambda=\pm 1$, as required. Since
\begin{equation}
C_\lambda
C_\mu=C_{\lambda+\mu}
\label{blah15}
\end{equation}
from consideration of the operator product
expansion, we can
evaluate $C_\lambda$ for all $\lambda\in\Lambda$ from its values on a
basis $\{\lambda_i\}$for the lattice. We may use these values to specify a
vector
$m\in\Lambda^\ast/2\Lambda$ such that $C_\lambda=(-1)^{\lambda\cdot m}$ for
the basis vectors, and hence for all vectors in the lattice from the
composition rule \reg{blah15}, by setting $m=\sum_im_i\mu_i$, where
$\{\mu_i\}$ is the dual basis to $\{\lambda_i\}$
and $m_i={1\over 2}(1-C_{\lambda_i})$.

Consideration of the action on the twisted sector must be divided into
the two cases $d=0\bmod 16$ and $d=8\bmod 16$. In the former case, the
ground state consists of spinors $\chi$\cite{DGMtwisted}. If we write
\begin{equation}
\theta\chi=S\chi
\end{equation}
for some matrix $S$, then
\begin{equation}
\theta V_0(|\lambda\rangle+|-\lambda\rangle)\chi=2^{-\lambda^2}
\theta\gamma_\lambda\chi
=2^{-\lambda^2}S\gamma_\lambda\chi\,,
\end{equation}
and
\begin{equation}
\theta V_0(|\lambda\rangle+|-\lambda\rangle)\chi=
V_0(\theta(|\lambda\rangle+|-\lambda\rangle))\theta\chi
=2^{-\lambda^2}C_\lambda\gamma_{R\lambda}S\chi\,.
\end{equation}
Hence
\begin{equation}
S\gamma_\lambda S^{-1}=C_\lambda\gamma_{R\lambda}\,,
\end{equation}
as required.

For $d=8\bmod 16$, we write
\begin{equation}
\theta c_{-1/2}^i\chi=c_{-1/2}^j{S^i}_j\chi\,,
\end{equation}
for some matrices ${S^i}_j$ acting on the spinor states upon which the
twisted sector is built. Then, for $i\neq j$ and dropping the
summation convention,
\begin{equation}
\theta V_0(a_{-1}^ia_{-1}^j|0\rangle)c_{-1/2}^i\chi={1\over 2}
\theta c_{-1/2}^j\chi
={1\over 2}\sum_kc_{-1/2}^k{S^j}_k\chi\,,
\end{equation}
and
\begin{eqnarray}
\theta V_0(a_{-1}^ia_{-1}^j|0\rangle)c_{-1/2}^i\chi&=&
\sum_{k,l,m}{S^{ij}}_{kl}V_0(a_{-1}^ka_{-1}^l|0\rangle)c_{-1/2}^m{S^i}_m\chi
\nonumber\\
&=&{1\over 2}\sum_{k,l}{S^{ij}}_{kl}\left({S^i}_kc_{-1/2}^l\chi
+{S^i}_lc_{-1/2}^k\chi\right)\,.
\end{eqnarray}
So
\begin{equation}
{S^j}_k=\sum_l\left({S^{ij}}_{kl}+{S^{ij}}_{lk}\right){S^i}_l\,.
\end{equation}
Set $\rho^l=\sum_k{R^l}_k{S^l}_k$ (still no summation convention).
Then, substituting in for ${S^{ij}}_{kl}$ into the above gives, using
$R\in{\rm O}(d)$, $\rho^i=\rho^j=S$, say, so that the equation becomes
\begin{equation}
{S^j}_k={R^j}_kS+\sum_l{R^i}_k{R^j}_l{S^i}_l\,.
\label{hmmm}
\end{equation}
If we denote $\sum_k{R^l}_k{S^m}_k$ by $\Theta^{lm}$, then \reg{hmmm}
gives us
\begin{equation}
\Theta^{mj}=\delta^{mj}S+\delta^{im}\Theta^{ji}\,.
\end{equation}
So for $m\neq j$, $i$, $\Theta^{mj}=0$. Since $i$ does not appear in
the definition of $\Theta$, then $\Theta^{mj}=0$ for $m\neq j$. Thus,
\reg{hmmm} reduces to
\begin{equation}
{S^j}_k={R^j}_kS+{R^i}_k\Theta^{ji}
={R^j}_kS\,,
\end{equation}
as required. Considering $\theta V_0(|\lambda\rangle +
|-\lambda\rangle)c_{-1/2}^i\chi$ as in the case of $d=0\bmod 16$ then
gives us $S\gamma_\lambda S^{-1}=C_\lambda\gamma_{R\lambda}$ again,
and the proof is complete.
\subsubsection{Continuous Automorphisms of Meromorphic CFT's}
In the way of an aside from the main discussion,
we may also consider the question of continuous automorphisms of
CFT's. It is a widely believed piece of folklore
in the subject that continuous symmetries are generated by the modes
of the vertex operators corresponding to states of conformal weight
one. However, there are clear counter-examples to this. For example,
consider a theory of the form $\Hil(\Lambda)_+$,
but take $\Lambda$ to be a $d$-dimensional zero
lattice. This theory clearly has an O$(d)$ symmetry group, but there
are no states of conformal weight one.

In general, let $\theta$ generate a continuous automorphism of a bosonic
hermitian
meromorphic CFT $\Hil$, {\em i.e.}
\begin{equation}
e^{a\theta}V(\psi,z)e^{-a\theta}=V(e^{a\theta}\psi,z)\,,
\end{equation}
for all $\psi\in\Hil$ and $a\in\re$, or more conveniently
\begin{equation}
[\theta,V(\psi,z)]=V(\theta\psi,z)\,.
\end{equation}
We shall assume that $\theta$ can be
written in terms of the vertex operators of the theory, and hence, by
duality and the fact that it must leave the Virasoro generators
invariant \cite{thesis}
(and hence the conformal weights), in the form
$\theta=V(\psi_\theta)_0$. We expand $\psi_\theta=\sum_{n\geq 1}\psi_n$, where
$\psi_n$ is a state of conformal weight $n$ (and we exclude the case
$n=0$ as it gives simply a trivial constant addition to $\theta$).
Noting that
\begin{equation}
V(L_{-1}\psi,z)=[L_{-1},V(\psi,z)]={d\over{dz}}V(\psi,z)\,,\label{deriv}
\label{s8}
\end{equation}
which gives a relation between the modes $V(L_{-1}\psi)_n$ and $V(\psi)_n$,
we see that we may redefine the $\{\psi_n\}$ if necessary so that they
are all quasi-primary.
Now, again from the invariance of the Virasoro generators under the
automorphism, we must have $[L_{\pm 1},\theta]=0$, and so
\begin{equation}
\sum_{n\geq 2}(n-1)V_{\pm 1}(\psi_n)=0\,.
\end{equation}

Let us assume that for $\psi$ quasi-primary
$V(\psi)_{-1}=0\Rightarrow\psi=\lambda|0\rangle$, for
some $\lambda\in\ce$ [we interpret $V(\phi)_m$ as $\sum_nV(\phi_n)_m$
where $\phi=\sum_n\phi_n$ is a decomposition of $\phi$ into states
$\phi_n$ of conformal weight $n$]. (The
equivalent statement for $V(\psi)_1$ clearly follows using the
hermitian structure of the theory.) Then we must have $\sum_{n\geq
2}(n-1)\psi_n=0$, and hence $\psi_n=0$ for $n\geq 2$ (since $\psi_n$ for
distinct $n$ are at distinct levels in $\Hil$, and thus linearly independent).
Thus, under this assumption, we see that any continuous symmetry must
be generated by the zero modes of the weight one states (if any) in
the theory.

Note that in the pathological theories based on a zero lattice
described above, we can construct a non-trivial state $\psi$ such that
$V(\psi)_{-1}=a^i_{-1}a^j_0$ (and further $\psi$ can be redefined if
necessary to be quasi-primary, using \reg{s8}),
which vanishes identically on the Hilbert space.

Let us consider the converse situation. Suppose we have a state
$\psi$ (consisting only of pieces of weights at least 2) such that
\begin{equation}
V_{-1}((L_0-1)\psi)\equiv [L_{-1},V_0(\psi)]=0\,.\label{ass}
\end{equation}
If we split $\psi$ as before into states
$\psi_n$ of weight $n$, then considering the operator product
expansion gives us
\begin{equation}
[V_0(\psi),V(\phi,w)]=V(V_0(\psi)\phi,w)+\sum_{n\geq
2}\sum_{0<m<n}V(V(\psi_n)_{-m}\left({n-1\atop m}\right)\phi,w)w^m\,.
\end{equation}
Then, using \reg{ass} and repeated applications of \reg{s8},
we see that the second term vanishes, and we
have that $V_0(\psi)$ generates a continuous automorphism of $\Hil$,
though of course not necessarily a non-trivial one.
Note that the ambiguity in defining the state $\psi$ afforded by
\reg{s8} (which we may use as above to make $\psi$ quasi-primary, for
the sake of definiteness) is irrelevant, for \reg{s8} gives
\begin{equation}
V_{-1}\left((L_{-1}+L_0-1)\psi\right)=0
\label{ddag}
\end{equation}
and
\begin{equation}
V_0\left({1\over L_0-1}(L_{-1}+L_0-1)\psi\right)=
V_0\left((L_{-1}+L_0){\psi\over L_0}\right)=0\,.
\end{equation}

Note that  another way of phrasing our condition on the
CFT is that $V_0(\psi)$ determines $\psi$ uniquely
up to states of conformal weight one if $\psi$
is quasi-primary.

[Suppose $\Hil$ is such that $V_0(\phi)=0$ and $\phi$ quasi-primary
implies that $\phi$ is of weight one (the same as the above statement
by linearity of the vertex operators in their arguments). Then suppose
that $V(\psi)_{-1}=0$. If $\psi$ is quasi-primary, the usual commutation
relations with $L_1$ give $V_0(L_0\psi)=0$, and so $L_0\psi$ is of weight
one (it is clearly quasi-primary),
{\em i.e.} $\psi=\lambda|0\rangle+\chi$, for some $\lambda\in\ce$
and $\chi$ of conformal weight one. But $V(\lambda|0\rangle+\chi)_{-1}
|0\rangle=\chi$, and so $V(\psi)_{-1}=0$ fixes $\chi=0$, and we have that
$V(\psi)_{-1}=0$ and $\psi$ quasi-primary implies that $\psi=\lambda
|0\rangle$ for some $\lambda\in\ce$.

Conversely, suppose that $V(\psi)_{-1}=0$ and $\psi$ quasi-primary
implies that $\psi=\lambda|0\rangle$ for some $\lambda\in\ce$. Then,
if $V_0(\phi)=0$, $[L_{-1},V_0(\phi)]=0$, {\em i.e.} $V((L_0-1)\phi)_{-1}
=0$. If $\phi$ is quasi-primary, then so is $(L_0-1)\phi$, and so we have
$(L_0-1)\phi=\lambda|0\rangle$ for some $\lambda\in\ce$, {\em i.e.}
$\phi=-\lambda|0\rangle+\chi$, for some state $\chi$ of weight one.
But $V_0(\phi)=0$, and so $\lambda=0$. Thus, we have the required
equivalence.]

We shall
say that a CFT is {\em deterministic} if it
satisfies this criterion. It remains to be proven even that the theories
$\Hil(\Lambda)$ and $\widetilde\Hil(\Lambda)$ are of this type,
though we know that the theory $\widetilde\Hil(\Lambda_L)$ (where
$\Lambda_L$ is the Leech lattice) has only discrete symmetries, from
\cite{FLMbook}.
\subsection{Application to the theories $\Hil(\Lambda)$}
Consider
projections by arbitrary automorphisms of finite order
of the Niemeier lattice theories
$\Hil(\Lambda)$.
Schellekens and Yankielowicz in \cite{SchellYank:curious}
have already observed that this is useful to consider, since
one of their two proposed new theories may be regarded
as a $\ze_2$-orbifold of the theory $\Hil({E_8}^3)$ induced by the
involution on the lattice which interchanges two of the $E_8$ factors and
shifts the third by a $D_8$ weight vector (although we feel that they
should
really consider a reflection on the third factor instead,
since the shift is {\em not} a lattice automorphism!).

Note that it seems to be known, at least to the
mathematicians\cite{Mythesis,Lepowsky}, how to construct the twisted vertex
operators corresponding to an arbitrary lattice automorphism, though
their results need to be corrected to compensate for the analogous
normal ordering problem to that described and solved in \cite{DGMtwisted}.
For an
involution, the techniques applied in \cite{DGMtwisted} may then be
used to construct the corresponding intertwining vertex operators
which enable one to unite the invariant theory and its representation
into a consistent CFT, while for higher order
automorphisms the techniques of \cite{PSMthird,PSMOhio} will be relevant.
The question of the consistency of such a
theory still has to be resolved by direct calculation though,
analogous to that carried out in \cite{DGMtwisted}. In any case, we
can certainly calculate the ground state energy and degeneracy of the
twisted sector, assuming consistency. Thus, use of the above
techniques may seem unnecessary, though they do provide a quicker
route to the answer and are really the only tool which can be used in
the more complex situation considered in the next subsection where no
explicit construction of the orbifold is yet known.

Note that not all automorphisms give rise to a consistent orbifold
theory. All cases of this which we will come across which are not on
Schellekens' list can be
eliminated simply by observing that the energy of the twisted sector
ground state is not in $\ze/p$. This is Vafa's ``level-matching
condition'' \cite{Vafa}, which is the condition for the absence of a
global phase anonmaly (assumed in our earlier discussions). What we do
can therefore be interpreted as either checking Vafa's conjecture, which was
made on the basis of assumptions similar to that we have made about
the transformation properties of the  constituent parts of the
orbifold partition function, by cross-checking with restrictions
provided by Schellekens' list, or checking our technique of
reference to Schellekens' list to determine consistency by reference
to this simple global phase test. The apparent validity of the check
against Schellekens' list as a means of establishing consistency of
the orbifold will thus be relied upon in the next section, where there
is no explicit construction known to us and hence no means of
calculating the twisted sector ground state energy directly.

Let us consider an example.
The glue code for the Niemeier lattice ${E_6}^4$ is the tetracode
$\C_4$\cite{ConSlo}.
This has an involution of the form $(\leftrightarrow 1 -1)$,
using the obvious notation.  If we try to take this as our involution,
it gives a twist invariant algebra ${E_6}^2C_4$, and the argument of
section \ref{applic} tells us that the number of weight one states in
the orbifold theory is $288-24\alpha$ (we can work out $\alpha$ if
required). The only possible theory from Schellekens' list of
sufficient rank would need $\alpha=5$ and have algebra $A_5C_5E_6$,
which is inconsistent with the twist invariant algebra.
The
ground state energy of the twisted sector is $6/16$
from the $-1$ on one $E_6$ plus $6/16$ from the interchange of the pair
of $E_6$'s, which is not half-integral, veri

Let us try replacing the $1$ in the specification of the automorphism
with an involution of $E_6$ (which leaves the glue fixed). We try a
one with invariant algebra $A_5A_1$\cite{Mythesis}, which we
can check leaves the glue unaffected. Thus, the twist
invariant algebra is $E_6A_5A_1C_4$, and we find the number of weight
one states to be $168-24\alpha$. This is consistent with the new
theory $A_5C_5E_{6,2}$, if we demonstrate that $\alpha=0$. Note that,
from \cite{Mythesis}, we see that the extra automorphism in this case
contributes $1/4$ to
the ground state energy in the twisted sector, making it integral (and
also incidentally implying that $\alpha=0$, since there are no weight
$1/2$ states). Also, note that the weight one state argument implies
that 16 new weight one states must come from the twisted sector, as
an explicit construction along the lines discussed in \cite{Mythesis}
would verify.

One may proceed systematically through the Niemeier lattices and the
involutions of each. We present ${E_8}^3$ as the simplest example.
Note that we merely consider the lattice involutions, whereas the
automorphism group of the CFT is extended by the presence of the
cocycles. The implication of the presence of additional involutions
due to these must also be taken into account. We will discuss this
briefly below.
We have three possibilities for contributions to the lattice
involution of ${E_8}^3$. We have transposition of a pair of the
$E_8$'s and also two involutions of the $E_8$ root lattice itself. We
refer to these as $\theta_1$ and $\theta_2$. $\theta_1$ is simply the
reflection, and gives a contribution of $1/2$ to the twisted sector
ground state energy with degeneracy 16 (see for example \cite{DGMtwisted}).
The corresponding invariant algebra is $D_8$. $\theta_2$ has invariant
subalgebra $E_7A_1$, and contributes $1/4$ to the energy with
degeneracy 2\cite{Mythesis}.

The results are tabulated in table \ref{e8inv}. We label the orbifold
by the algebra, though of course we only know the corresponding theory
to (exist and) be unique in the rank 24 case\cite{thesis}.
We see from applying our above arguments to the calculation of
$\alpha$ that we may interpret the transposition of two components of
the lattice as contributing $1/2$ to the energy of the ground state in
the twisted sector with unit degeneracy (independent of whether we
attach another automorphism). Note that theory 3 would give
$N=432-24\alpha$, with a subalgebra $E_8E_7A_1$. We can eliminate this
without even bothering to calculate $\alpha$ as we see that there is
no suitable theory in Schellekens' list. Again, we can also
eliminate it immediately since it would require a ground state energy
of $3/4$ in the twisted sector.

Note that in some places we use known results for the
reflection twist\cite{DGMtrialsumm}, {\em i.e.} for the involutions 8
and 11, whereas for
1, 10, 13 and 16 we do not have sufficient knowledge from these simple
calculations to identify the CFT completely.
Thus, we obtain the four distinct theories $\Hil(D_{10}{E_7}^2)$,
$\Hil({E_8}^3)$, $\Hil(E_8D_{16})$ and $\Hil({D_8}^3)$ and at least
one theory with algebra $E_{8,2}B_{8,1}$ by orbifolding $\Hil({E_8}^3)$
with respect
to the lattice induced involutions.

\begin{table}[htb]
\begin{center}
\begin{tabular}{|c|c|c|c|}\hline
Label & Involution & Orbifold & $\alpha$ \\ \hline
1 & $\leftrightarrow\cdot$ & ${E_8}^3$ or $E_8D_{16}$ & 1 \\
2 & $\leftrightarrow\theta_1$ & $E_{8,2}B_{8,1}$ & 0 \\
3 & $\leftrightarrow\theta_1$ & incorrect vacuum energy & $-$ \\
4 & $\theta_1\theta_2\theta_2$ & $D_{10,1}{E_{7,1}}^2$ & 0 \\
5 & $\theta_1\theta_1\theta_2$ & incorrect vacuum energy & $-$ \\
6 & $\theta_1\theta_1\theta_1$ & ${D_8}^3$ & 0 \\
7 & $\theta_2\theta_2\theta_2$ & incorrect vacuum energy & $-$ \\
8 & $\cdot\ \theta_1\theta_1$ & $E_8D_{16}$ & 0 \\
9 & $\cdot\ \theta_1\theta_2$ & incorrect vacuum energy & $-$ \\
10 & $\cdot\ \theta_2\theta_2$ & ${E_8}^3$ or $E_8D_{16}$ & 4 \\
11 & $\cdot\cdot\theta_1$ & ${E_8}^3$ & 16  \\
12 & $\cdot\cdot\theta_2$ & incorrect vacuum energy & $-$ \\
13 & $(\theta_1\leftrightarrow\theta_1)\ \cdot$ & ${E_8}^3$ or
$E_8D_{16}$ & 1 \\
14 & $(\theta_1\leftrightarrow\theta_1)\theta_1$ & $E_{8,2}B_{8,1}$ & 0 \\
15 & $(\theta_1\leftrightarrow\theta_1)\theta_2$ & incorrect vacuum
energy & $-$ \\
16 & $(\theta_2\leftrightarrow\theta_2)\ \cdot$ & ${E_8}^3$ or
$E_8D_{16}$ & 1 \\
17 & $(\theta_2\leftrightarrow\theta_2)\theta_1$ & $E_{8,2}B_{8,1}$ & 0 \\
18 & $(\theta_2\leftrightarrow\theta_2)\theta_2$ & incorrect vacuum
energy & $-$ \\ \hline
\end{tabular}
\caption{Involutions of ${E_8}^3$ and the corresponding orbifolds of
$\Hil({E_8}^3)$.}
\label{e8inv}
\end{center}
\end{table}

We may briefly consider the extension of the automorphism group of the
lattice due to the presence of the cocycles. We have automorphisms
$u a_n^j u^{-1}=R_{ij} a_n^i$,
$u|\lambda\rangle=(-1)^{\lambda\cdot\mu}|R\lambda\rangle$, $R\in{\rm
Aut}\ (\Lambda)$, $\mu\in\Lambda/2\Lambda$. Take the simple case
$R=1$. We use the construction for the root lattice $E_8$ given in the
appendix of Myhill's thesis. Then there are two inequivalent choices
for $\mu$, {\em i.e.} $e_1+e_2$ and ${1\over 2}\sum_{i=1}^8e_i$. These
are both found to give 136 invariant states, {\em i.e.} they behave
in the same way as $\theta_2$ on each component.

We may check the analysis of section \ref{qwer}
by computing $C_0$ in the case of a third
order
no-fixed-point automorphism of a $c=24$ lattice theory
$\Hil(\Lambda)$\cite{thesis}. The theory can be taken to be built up by two
sets of
oscillators $a^i_n$ and $\bar a^i_n$, $1\leq i\leq 12$, $n\in\ze$,
from momentum states $|\lambda\rangle$ with $\lambda\in\Lambda_{\cal
E}$, an Eisenstein lattice whose corresponding $\ze$-lattice (given by
splitting into real and imaginary parts and scaling by $\sqrt 2$) is
$\Lambda$.
The oscillators satisfy the relations
\begin{equation}
\left[a^i_m,a^j_n\right]=\left[\bar a^i_m,\bar a^j_n\right]=0\,;
\quad \left[a^i_m,\bar a^j_n\right]=m\delta_{m,-n}\delta^{ij}
\end{equation}
for $1\leq i,j\leq 12$, $m$, $n\in\ze$,
and $a^i_n|\lambda\rangle=\bar a^i_n|\lambda\rangle=0$ for $n>0$,
$\lambda\in\Lambda_{\cal E}$, with $a^i_0|\lambda\rangle=\lambda^i
|\lambda\rangle$ and $\bar a^i_0|\bar\lambda\rangle=\bar\lambda^i
|\lambda\rangle$. The automorphism $\theta$ is given by $\theta
a^i_n\theta^{-1}=\omega a^i_n$, $\theta\bar a^i_n\theta^{-1}=\bar
\omega\bar a^i_n$ and $\theta|\lambda\rangle=|\omega\lambda\rangle$,
where $\omega=e^{2\pi i/3}$. It is now trivial to work out the
invariant states at each level (the conformal weight of the state
$\prod_{a=1}^M\prod_{b=1}^Na^{i_a}_{-m_a}\bar
a^{j_b}_{-n_b}|\lambda\rangle$ is given by
$\sum_{a=1}^Mm_a+\sum_{b=1}^N n_b+\lambda\cdot\bar\lambda$), and we
find that
\begin{equation}
P_1={1\over 3}|\Lambda(2)|\quad,
P_2={1\over 3}|\Lambda(4)|+8|\Lambda(2)|+144\quad,
P_3={1\over 3}|\Lambda(6)|+8|\Lambda(4)|+108|\Lambda(2)|+1016\,,
\end{equation}
where $|\Lambda(n)|$ is the number of vectors of length squared $n$ in
the lattice $\Lambda$.
We know that, for an even self-dual lattice, the theta function is
given by
\begin{equation}
\Theta_\Lambda(\tau)=(J(\tau)+D)\eta(\tau)^{24}\,,
\end{equation}
for some constant $D$,
and so we find that $P_1={1\over 3}D-8={1\over 3}|\Lambda(2)|$,
$P_2=65664$ and $P_3=7164536$. This gives us $C_0={4\over
3}|\Lambda(2)|+24$, as is clear from the explicit construction as the
sum of the number of weight one states in $\Hil(\Lambda)$ and the
corresponding $\ze_3$-orbifold.

Note that it can be instructive to evaluate $C_0$ in terms of the $M_i$, which
we
know to be non-negative elements of $\ze/p$ (if the orbifold construction is
consistent).
We obtain
\begin{equation}
C_0=4P_1+12(2-9M_1-3M_2)
\label{blah}
\end{equation}
for $p=3$, for example. We then obtain an upper bound on the number of
weight one states in the orbifold theory, a lower bound being given
trivially by the number of weight one states in the common sub-theory.

As an example to illustrate that it is worthwhile to consider these
higher order twisted theories, let us demonstrate the existence of a
new theory at $c=24$. Consider the theory $\Hil({E_6}^4)$, where
${E_6}^4$ indicates the Niemeier lattice with the corresponding root
system. Consider the automorphism of this lattice given by cycling
three of the $E_6$ factors and acting with the third order
automorphism $\theta$
of $E_6$ on the fourth which leaves an invariant algebra
${A_2}^3$\cite{Mythesis}. It can be checked that this is actually an
automorphism of the Niemeier lattice. From the twisted sector ground
state energy corresponding to $\theta$ given in \cite{Mythesis}, we
see that the twisted sectors must have lowest conformal weight 1.
Thus, $M_1=M_2=0$, and so we can evaluate $C_0$ directly from
\reg{blah}. We thus find that there are 120 weight one states in the
orbifold CFT, assuming it to be consistent, and we
have the invariant algebra $E_6{A_2}^3$. Comparing with the list of
possibilities derived in \cite{SchellComplete}, we see that we must
have algebra $E_{6,3}(G_{2,1})^3$, since this is the only possibility
for 120 weight one states, and moreover is consistent with the
invariant algebra, suggesting that the orbifold is in fact consistent,
though this still must be checked explicitly of course.
\subsection{Application to the theories $\widetilde\Hil(\Lambda)$}
As we remarked above, the application to theories $\Hil(\Lambda)$ is
in some sense trivial, though it has provided us with some new
theories and verified our technique for checking consistency.
Let us now consider projecting out by a
lattice induced involution the theories $\widetilde\Hil(\Lambda)$ for
$\Lambda$ self-dual.
This could contain the analogue of the ternary code $\ze$-lattice
constructions of \cite{PSMcodes} to which we have referred previously,
since their structure seems to indicate a $\ze_2\times\ze_2$ orbifold.
This, together with the fact that these are the only theories other
than $\Hil(\Lambda)$ known to exist at $c=24$, provides us with
sufficient motivation for such an investigation.
There are two cases to consider.

The first is that in which
$\widetilde\Hil(\Lambda)\cong\Hil(\Lambda')$  for some even
self-dual lattice $\Lambda'$
(of which there are 9
cases for $c=24$, {\em i.e.} one for each doubly-even self-dual binary linear
code in 24 dimensions\cite{DGMtrialsumm}).
In this case, it may be argued that the lattice induced
automorphism is not lattice induced from the point of view of the
theory $\Hil(\Lambda')$, and hence that the above arguments would yield
non-trivial information about a new orbifold theory.
However,
as we demonstrated in section \ref{flobtip}, this is not the case.

The second possibility is that $\widetilde\Hil(\Lambda)\not\cong
\Hil(\Lambda')$ for any $\Lambda'$ (15 instances in $c=24$).
This means that we are in a non-trivial situation, {\em
i.e.} the explicit twisted vertex operator construction of \cite{Mythesis} is
not valid.

Consider a lattice induced automorphism $\theta$ of
$\widetilde\Hil(\Lambda)$.
This is given by \reg{amv}, and we may take $v_{R,S}(\lambda)=(-1)^{
m\cdot\lambda}$, $m\in\Lambda/2\Lambda$.
We can easily write an explicit
expression for $T_\theta(\tau)\equiv{\theta\lower 6pt \hbox{$
{\square\atop\displaystyle 1}$}}(\tau)$\cite{FLMbook}, {\em i.e.}
\begin{equation}
T_\theta(\tau)={1\over
2}\left[{\Lambda_R^m(\tau)\over{\eta_R(\tau)}}+
{\Lambda_{-R}^m(\tau)\over{\eta_{-R}(\tau)}}\right]
+{1\over 2}{\rm
Tr}S\left[{\eta_R(\tau)\over{\eta_R({\tau\over 2})}}
+(\tau\mapsto\tau+1)\right]\,,
\end{equation}
where $\Lambda^m_R(\tau)=\sum_{\{\lambda\in\Lambda:R\lambda=
\lambda\}}(-1)^{m\cdot\lambda}q^{{1\over 2}\lambda^2}$ and
$\eta_R(\tau)=\prod_{k|N}\eta(k\tau)^{a_k}$, with $R$ of
order $N$ and
${\rm det}(x-R)=\prod_{k|N}(x^k-1)^{a_k}$.

 From our preceding arguments, in order to evaluate the number of
weight one states in the orbifold of $\widetilde\Hil(\Lambda)$ with
respect to $\theta$ we need the number of invariant states at levels up
to $p$ for $N=p=2$, 3, 5, 7. Clearly, for $p>2$ a prime, $\Lambda^m_{
R^i}(\tau)=\Lambda^m_R(\tau)$,
$\Lambda^m_{
-R^i}(\tau)\equiv 1$ and $\eta_{R^i}(\tau)=\eta_R(\tau)$
 for $1\leq i\leq p-1$. Hence the partition function of the invariant
theory is given by
\begin{equation}
S_\theta(\tau)={1\over p}\left(J(\tau)+{1\over
2}|\Lambda(2)|+(p-1)T_\theta(\tau)\right)\,,
\end{equation}
using \reg{gibbon}, where
\begin{equation}
T_\theta(\tau)={1\over
2}\left[{\Lambda_R^m(\tau)\over{\eta_R(\tau)}}+
{1\over{\eta_{-R}(\tau)}}\right]
+{1\over 2}{\rm
Tr}S\left[{\eta_R(\tau)\over{\eta_R({\tau\over 2})}}
+(\tau\mapsto\tau+1)\right]\,.
\label{kebab}
\end{equation}
Now, we also have $S_\theta(\tau)$ in the form \reg{thingy}. Comparing
the two forms will allow us, in certain cases, to evaluate ${\rm Tr}S$
in terms of $a_k$ and $\Lambda^m_R$, and our conjecture would
then be that evaluation of this trace by the more conventional means
arising from gamma matrix theory will provide a further consistency
check,
analogous to Vafa's condition in the previous section.

In the case $p=2$, we then compare
\begin{eqnarray}
S_\theta(\tau)-{1\over 2}J(\tau)&=&{1\over 2}T_\theta(\tau)\nonumber\\
&=&\alpha f_2(\tau)+\beta g_2(\tau)\,,
\end{eqnarray}
where $\alpha+\beta={1\over 2}$ (ignoring additive constants).
The second form implies that $P_3=24
P_2+ 8379936$, while the first gives us $P_2={1\over 2}{\rm Tr}R\,{\rm
Tr}S+\cdots$ and $P_3=\left({1\over{12}}({\rm Tr}R)^3+{20\over 3}{\rm
Tr}R\right){\rm Tr}S+\cdots$ (${\rm Tr}R=a_1$). Thus, we are able to solve for
${\rm
Tr}S$ except in the cases ${\rm Tr}R=0$, $\pm8$. In order to
demonstrate that consideration of higher order terms cannot allow us to
solve for ${\rm Tr}S$ in these cases, we must verify that in such
cases
\begin{equation}
f_2(\tau)-g_2(\tau)\propto\left[{\eta_R(\tau)\over{\eta_R({\tau\over
2})}}+(\tau\mapsto\tau+1)\right]\,.
\end{equation}
[Note, however, that even if we can solve for ${\rm Tr}S$ then this is
still no guarantee that the two forms for $S_\theta(\tau)$ are
identical.]
For ${\rm Tr}R=0$, the term proportional to ${\rm Tr}S$ in \reg{kebab}
vanishes identically (and so there is no need to know ${\rm Tr}S$ in
order to evaluate $S_\theta(\tau)$), while for ${\rm Tr}R=\pm 8$ it is
\begin{equation}
\pm\left({\eta(2\tau)^8\over{\eta({\tau\over 2})^8}}-{\eta(2\tau)^{16}
\eta({\tau\over 2})^8\over{\eta(\tau)^{24}}}\right)\,.
\end{equation}
But, from \cite{Tuite:moon},
\begin{equation}
f_2(\tau)-g_2(\tau)=\left({\eta(\tau)\over{\eta(2\tau)}}\right)^{24}-
{(\theta_3(\tau)\theta_4(\tau))^8+2^{-4}\theta_2(\tau)^{16}\over
{\eta(\tau)^8\eta(2\tau)^8}}\,.
\end{equation}
Now, from the product form for the theta functions\cite{BRB},
\begin{equation}
\theta_2(\tau)=2{\eta(2\tau)^2\over{\eta(\tau)}}\quad,
\theta_3(\tau)={\eta(\tau)^5\over{\eta(2\tau)^2\eta({\tau\over
2})^2}}\quad,
\theta_4(\tau)={\eta({\tau\over 2})^2\over{\eta(\tau)}}\,.
\end{equation}
We then see that we will be unable to infer ${\rm Tr}S$ if
\begin{equation}
\eta(\tau)^{24}-\eta(2\tau)^8\eta({\tau\over
2})^{16}=16\eta(2\tau)^{16}
\eta({\tau\over 2})^8\,,
\end{equation}
fixing the constant of proportionality from the first term in the
expansion.
This is easily seen to be equivalent to the Jacobi triple
identity
${\theta_2}^4+{\theta_4}^4={\theta_3}^4$, and so we have the desired result.
Note that a necessary (and we would conjecture a sufficient) condition
for $\Hil(\lambda)_\theta$ to be consistent is that ${\rm Tr}R=\pm
24$, $\pm8$ (from requiring that the conformal weights in the twisted
sector be half-integral), and we could posit that
$\widetilde\Hil(\Lambda)_\theta$ is also only consistent for these
values (because $\widetilde\Hil(\Lambda)=\Hil(\Lambda)_{-1}$ and the
automorphisms $-1$ and $R$ commute). Thus, we would expect that it
is impossible to deduce ${\rm Tr}S$ and hence the partition function
of $\widetilde\Hil(\Lambda)_\theta$ merely by consideration of the
modular transformation properties, and we must consider the gamma
matrix theory approach for this case.

However, $p=2$ is usually a very special case, particularly in this
instance where a second order automorphism is already involved in
the definition of $\widetilde\Hil(\Lambda)$. So, we consider the
case $p=3$.

The ansatz \reg{thingy} gives $N_4=8182-198N_2+24N_3$, and then the
explicit form \reg{kebab} for $T_\theta$ (noting that there are no
fixed points of $-R$, and that the fixed points and sets of
eigenvalues of $R$ and
$R^2$ coincide so that $T_\theta=T_{\theta^2}$)
gives, denoting ${\rm Tr}R$ by $R$ and ${\rm Tr}S$ by $S$,
\begin{eqnarray}
\left(
{1\over{120}}R^5+{1\over{12}}R^4-{27\over 8}R^3-{29\over
4}R^2+{896\over 5}R-192
\right) S=
-{1\over{24}}R^4-{1\over 4}R^3-{613\over 8}R^2
-{431\over 4}R
\nonumber\\
+12273-
{1\over 2}f_4+(12-{1\over 2}R)f_3-({1\over 4}R(R-45)-99)f_2
-({1\over{12}}R^3+{21\over 4}R^2-{163\over 2}R-4)f_1\,.
\label{t3}
\end{eqnarray}
We note that the coefficient of $S$ does not vanish for any of the
possible values of $R$, and hence that we may solve for $S$ as required.

As a simple example (and a check on the manipulations), let us consider
a third order NFPA, for example arising from the Eisenstein
formulation discussed above. In this case, we have no eigenvalues 1,
and so $a_1+a_3=0$. But $a_1+3a_3=24$, and hence $R=a_1=-12$. Also
$f_i=0$ clearly, and the above relation gives us $S=1$.
Thus, this
suggests that the coefficient of $S$ and the terms on the right hand
side not involving $f_i$ in \reg{t3} are correct.
Also, substituting
this into the expression for $T_\theta(\tau)$ gives us $N_2=36$,
$N_3=-{152\over 3}$. But $N_1={1\over 6}|\Lambda(2)|$ (since we see
that from the explicit construction that the invariant states of
weight one are given by $|\lambda\rangle+|-\lambda\rangle+
|\omega\lambda\rangle+|-\omega\lambda\rangle+|\bar\omega\lambda\rangle
+|-\bar\omega\lambda\rangle$ for $\lambda\cdot\bar\lambda=1$),
and hence $C_0={2\over 3}|\Lambda(2)|+6$, from \reg{star}. This tells
us that the number of weight one states in the orbifold of
$\widetilde\Hil(\Lambda)$ with respect to the third order automorphism
is ${1\over 6}|\Lambda(2)|+6$, of which ${1\over 6}|\Lambda(2)|$ come
from the invariant sector of $\widetilde\Hil(\Lambda)$. However, from
\cite{thesis} we have that the only Niemeier lattices admitting a
third order NFPA are ${E_8}^3$, ${A_2}^{12}$, ${E_6}^4$ and
$\Lambda_{24}$. The number of weight one states in the
corresponding orbifolds of $\widetilde\Hil(\Lambda)$ would be 126, 18,
54 and 6 respectively, none of which are allowed according to
Schellekens' classification. Therefore we conclude that the orbifolds
are not consistent. Thus, we have also shown that, though the above
calculation may produce a reasonable value for $S$, this is no
guarantee of the existence of the orbifold.

[Note however that if we first orbifold $\Hil(\Lambda)$ by the third
order NFPA and then orbifold subsequently by the reflection twist, we
will obtain ${1\over 6}|\Lambda(2)|+24(1-\alpha)$ for the number of
weight one states on the resulting potential orbifold theory. This is
distinct from the above ($\alpha\in\ze$), disproving the naive
expectation that
commuting automorphisms of the theory $\Hil(\Lambda)$ commute as
orbifold constructions, though it must be remembered that the orbifolds here
are
inconsistent.]

The cases $p=5$ and $7$ are similar.
\section{Conclusions}
We started from results for CFT's by Schellekens analogous to
some of those proved for the
Niemeier lattices by Venkov. The notion of
complementary representations for involutions
and in general an ansatz for the generalised Thompson series
in terms of known hauptmoduls,
coupled with the restrictions on
Kac-Moody algebras in the theories derived by Schellekens, then enabled
us, for the first time, to investigate with some degree of confidence
orbifold theories for which no explicit construction is known.

It remains to present a more systematic survey of the orbifolds of the
theories $\widetilde\Hil(\Lambda)$ for arbitrary automorphisms, though
we
have shown that the full classification of the orbifolds of the
theories
$\Hil(\Lambda)$ is now merely a matter of straightforward calculation
(using the techniques demonsrated) from the known Niemeier lattice
automorphisms.
This
will require explicit knowledge of the action of the involution on the
twisted sector ground states, as discussed in \cite{GNOS}, together of
course with the determination of the full automorphism group of these theories.
Also, we need to understand the conditions under which the orbifolds
are consistent.

Finally, the obvious hope for the future is to complete the analogue
of Venkov's results for CFT's. Venkov showed that,
for each of the
possible semi-simple algebras which his conditions selected, that there was
one and only one corresponding even self-dual lattice. His approach
was based on coding theory via the idea of glue codes. (Any relation
to the code structures investigated in \cite{DGMtrialsumm,PSMcodes,thesis} is
as yet not understood.) However, the analogue cannot be exact, for
Schellekens has demonstrated that for some of the algebras listed in
\cite{Schell:Venkov} there is no modular invariant combination of
Kac-Moody algebras, and hence there can be no consistent orbifold
theory. Nevertheless, it might be argued that, where a theory does exist,
it is unique. There are, as yet, no counterexamples to such a claim.
\section{Acknowledgements}
I would like to thank Peter Goddard for useful conversations and
Gonville and Caius College, Cambridge for funding of this work via a
Research Fellowship.
\appendix
\section{Linearity of $\mu(\lambda)$}
\label{app11}
In this appendix, we prove the linearity of the map
$\lambda\mapsto\mu(\lambda)$ for the automorphism of the theory
$\widetilde\Hil(\Lambda)$ considered in section \ref{flobtip}.

We first note that \reg{del11} may be extended to the whole of
$\re^d$, rather than just restricted to points on the lattice
$\Lambda$. This is because this equation is of
polynomial type, and its vanishing on the points of the lattice
must therefore
imply that it vanishes identically. Hence, for any $x\in\re^d$,
${S^{ij}}_{kl}x^kx^l$ is of the form $y(x)^iy(x)^j$ for some $y(x)\in\re^d$,
with $y(x)^2=x^2$.
Similarly, we can extend \reg{del12} to all points in $\re^d$, {\em
i.e.} $y(x+x')=\pm(y(x)\pm y(x'))$.

So, if we can define $y(x)$ locally such that $y(x+x')=y(x)+y(x')$,
then it will hold globally by continuity, provided our locally
continuous definition of $y(x)$ is globally well-defined.

Fix $1\leq i_0$, $j_0\leq d$. Then we may write
\begin{equation}
{y^i(x)\over y^{i_0}(x)}={{S^{ij_0}}_{kl}x^kx^l\over
{S^{i_0j_0}}_{mn}x^mx^n}\,.
\end{equation}
But $y(x)^2=x^2$, and so we write
\begin{equation}
y^i(x)=\pm{|x|{S^{ij_0}}_{kl}x^kx^l\over\sqrt{{S^{pj_0}}_{mn}x^mx^n
{S^{pj_0}}_{rs}x^rx^s
}}\,,
\label{frtg}
\end{equation}
(with no summation over $j_0$).

The map $x\mapsto y(x)$ is defined up to a sign. Considering the
restriction of this to $S^{d-1}$, {\em i.e.} we exclude the origin
(the only point at which $y(x)$ vanishes) since we need only consider
the topology of the setup,
we have a map $S^{d-1}\rightarrow
S^{d-1}/\ze_2$. We see that global consistency of the map \reg{frtg}
with a particular choice of sign
is equivalent to finding a lift of this line bundle to a map
$S^{d-1}\rightarrow S^{d-1}$. But $\pi_1(S^{d-1})$ is trivial for
$d>2$, and so there can be no global ambiguity in sign, since any loop
over which a potential sign change occurs can be shrunk continuously
to a point.

Although throughout this paper we are primarily concerned with
self-dual lattices, for the sake of completeness let us establish the
result in one and two dimensions also, so that the conclusion of section
\ref{flobtip} holds for $\Lambda$ not necessarily self-dual. Though we
know from \cite{DGMtwisted} that $\widetilde\Hil(\Lambda)$ is only
consistent as a CFT for $d$ a multiple of 8 (and
$\sqrt 2\Lambda^\ast$ even), the result for general $d$ and even
$\Lambda$ will be of use in considering the symmetries of (not
necessarily meromorphic) representations of $\Hil_+(\Lambda)$.

For $d=1$, it is trivial to make sign choices such that
$\mu(n\lambda)=n\mu(\lambda)$. For example, $\mu(2\lambda)=
\pm(\mu(\lambda)\pm\mu(\lambda))$. But $\mu(2\lambda)^2=(2\lambda)^2$.
So we must have $\mu(2\lambda)=\pm 2\mu(\lambda)$, and a choice of
sign gives the required result. Similarly, the remaining sign choices
are forced upon us, and we see trivially that such choices are
mutually consistent.

Let $\{\lambda_1$, $\lambda_2\}$ be a basis for $\Lambda$ in the case
$d=2$. By considering the action of $\theta$ on the operator product
expansion of $V(|\lambda_1\rangle+|-\lambda_1\rangle,z)
V(|\lambda_2\rangle+|-\lambda_2\rangle,w)$, we see that
$\mu(\lambda_1)\cdot\mu(\lambda_2)=\pm\lambda_1\cdot\lambda_2$. Thus,
$\mu(\lambda_1)$ and $\mu(\lambda_2)$ are linearly independent.
As in the $d=1$ case, we can choose signs such
that $\mu(n\lambda_1)=n\mu(\lambda_1)$, and likewise for
$\mu(n\lambda_2)$.
Now, suppose that we have chosen signs such that
$\mu(m\lambda_1+n\lambda_2)=m\mu(\lambda_1)+n\mu(\lambda_2)$ for all
$|m|+|n|\leq N$. Then for $|m|+|n|=N+1$, \reg{del12} gives
\begin{eqnarray}
\mu(m\lambda_1+n\lambda_2)&=&\pm(\sgm\mu(\lambda_1)\pm\mu((m-\sgm)\lambda_1+n\lambda_2))\nonumber\\
&=&\pm\sgm(\mu(\lambda_1)\pm((m-\sgm)\mu(\lambda_1)+n\mu(\lambda_2)))
\end{eqnarray}
and
\begin{eqnarray}
\mu(m\lambda_1+n\lambda_2)&=&\pm(\mu(m\lambda_1+(n-\sgn)\lambda_2)\pm\sgn\mu(\lambda_2)\nonumber\\
&=&\pm(m\mu(\lambda_1)+(n-\sgn)\mu(\lambda_2)\pm\sgn\mu(\lambda_2))\,.
\end{eqnarray}
Consistency of these two equations requires
\begin{equation}
\mu(m\lambda_1+n\lambda_2)=\pm(m\mu(\lambda_1)+n\mu(\lambda_2))\,,
\end{equation}
and we choose the sign as appropriate to obtain the required result by
induction.

Note that a trivial extension of this argument will work for arbitrary
values of $d$, though the argument given in the general case is more
appealing.

\end{document}